\documentclass{article}
\usepackage{nature,lscape,graphicx,fullpage,doublespace}

\begin{document}
\begin{spacing}{1.5}

\title{Is the Outer Solar System Chaotic?}
\author{Wayne B. Hayes}
\maketitle

{\centerline{Wayne Hayes}
 \centerline{Computer Science Department}
 \centerline{University of California, Irvine}
 \centerline{Irvine, California 92697-3435}
 \centerline{\tt wayne@ics.uci.edu}
}

\begin{quote}{\normalfont\fontsize{8}{10}\selectfont
{\bfseries One-sentence summary:}  Current observational uncertainty
in the positions of the Jovian planets precludes deciding whether or not
the outer Solar System is chaotic.
\par}
\end{quote}

\begin{quote}{\normalfont\fontsize{8}{10}\selectfont
{\bfseries 100 word technical summary:}
The existence of chaos in the system of Jovian planets has been in question
for the past 15 years. Various investigators have found Lyapunov
times ranging from about 5 millions years upwards to infinity, with no
clear reason for the discrepancy.
In this paper, we resolve the issue.  The position of the outer planets
is known to only a few parts in 10 million.  We show that, within that
observational uncertainty, there exist Lyapunov
timescales in the full range listed above.  Thus, the ``true'' Lyapunov
timescale of the outer Solar System cannot be resolved using current
observations.
\par}
\end{quote}

\begin{quote}{\normalfont\fontsize{8}{10}\selectfont
{\bfseries 100 word summary for general public:}
The orbits of the inner planets (Mercury, Venus, Earth, and Mars) are
practically stable in the sense that none of them will collide or be
ejected from the Solar System for the next few billion years.  However,
their orbits are chaotic in the sense that we cannot predict their
angular positions {\em within} those stable orbits for more than about
20 million years.  The picture is less clear for the outer planets
(Jupiter, Saturn, Uranus and Neptune).  Again their orbits are practically
stable, but it is not known for how long we can accurately predict their positions
within those orbits.
\par}
\end{quote}

{\bf
The existence of chaos among the Jovian planets is a contested issue.
There exists apparently unassailable evidence both that the outer Solar
System is chaotic\cite{SussmanWisdom92,MurrayHolman99}, and that it
is not\cite{Laskar94,GrazierNewmanKaulaHyman99,VaradiRunnegarGhil03}.
The discrepancy is particularly disturbing given that computed chaos is
sometimes due to numerical artifacts\cite{HerbstAblowitz89,WisdomHolman92}.
In this paper we discount the possibility of numerical artifacts, and demonstrate
that the discrepancy seen between various investigators is {\em real}.
It is caused by observational uncertainty in the orbital positions of the
Jovian planets, which is currently a few parts in 10 million.  Within that
observational uncertainty, there exist clearly chaotic trajectories
with complex structure and Lyapunov times ranging from 2 million years
to 230 million years, as well as trajectories that show no evidence of
chaos over 1Gy timescales.  Determining the true Lyapunov time of the outer
Solar System will require a more accurate observational determination
of the orbits of the Jovian planets.  Fully understanding the nature
and consequences of the chaos may require further theoretical development.
}

The Solar System
is known to be ``practically stable'', in the sense that none
of the 9 planets is likely to suffer mutual collisions, or be
ejected from the Solar System, over the next several billion
years\cite{Laskar94,Laskar96,Laskar97,ItoTanikawa02}.
The motion of Pluto is chaotic with a Lyapunov time of 10--20 million
years\cite{SussmanWisdom88,WisdomHolman91,SussmanWisdom92}, while the
inner Solar System has a Lyapunov time of about 4--5 million years\cite{Laskar89,Laskar90,SussmanWisdom92}.
Pluto's chaos is caused by the overlap of two-body
resonances\cite{MalhotraWilliams97}.  The cause of chaos in the inner
Solar System is not yet fully understood\cite{MurrayHolman01},
although secular resonances likely play a role\cite{Laskar90,Laskar94}.

The existence of chaos amongst the Jovian planets
(Jupiter, Saturn, Uranus, and Neptune) is less certain.
On one hand, Laskar\cite{Laskar89,Laskar90,Laskar94} found that
the orbits of the outer planets appeared non-chaotic.  This is
consistent with the knowledge that there are no {\em two}-body resonances
among the outer planets\cite{MurrayHolman99}, but nonetheless
Laskar's ``averaged'' integrations could not detect mean-motion resonances,
even if they existed.
On the other hand, chaos {\em was} found in a non-averaged, full
integration of the 9 planets by Sussman and Wisdom\cite{SussmanWisdom92}.
However, their computed Lyapunov time varied with simulation
parameters for reasons which were (at the time) unknown.

Two obvious but mutually-exclusive explanations can be offered
to explain why some investigators find chaos while others do not.
First, Sussman and Wisdom's widely varying Lyapunov time might be
the result of numerical artifacts, rather than physical effects.
This hypothesis was supported at the time by the lack of an
explanation for chaos in the outer Solar System.
A second plausible explanation is that since Laskar's averaged equations
do not model planet motion, Sussman and Wisdom were observing real chaos,
caused by something other than overlap of two-body resonances.

Unfortunately, {\em both} explanations have since been
convincingly offered, and neither has been disproved to date.
One one hand, chaos {\em can} be a numerical artifact\cite{HerbstAblowitz89},
even in an n-body integration\cite{WisdomHolman92,NewmanEtAlDDA2000}.
Furthermore, numerous careful integrations of the outer Solar System
have been performed specifically
to ensure the accuracy and convergence of the numerical
results\cite{GrazierNewmanKaulaHyman99,VaradiRunnegarGhil03},
and these give a clear indication of {\em no chaos}.
On the other hand, an explanation for the chaos in
terms of {\em three}-body mean-motion resonances has been offered
by Murray and Holman\cite{MurrayHolman99}.
Furthermore, Guzzo\cite{Guzzo05} has performed carefully-tested integrations
detecting a web of three-body resonances, precisely where Murray and
Holman's theory predicts them to be.  Finally, those who have found
chaos also appear to perform reasonable convergence tests, making it
unlikely that the chaos they have found is a numerical artifact.  So
we have a quandary: there is apparently unassailable evidence on two
sides, demonstrating both that the outer Solar System is chaotic,
and that it is not.

The observational uncertainty in the position of the outer planets is a few parts in
ten million\cite{Standish98-DE405}.
The resolution of the above paradox is simple: within the
observational uncertainty, there exist both chaotic and non-chaotic solutions.
This is also consistent with Murray and Holman's theory, since it is known
from the study of simpler systems that chaotic ``zones'' can contain both
chaotic and regular-looking trajectories, densely packed amongst each other,
with the chaotic ones having widely varying Lyapunov times \cite{CorlessWhatGood94}.

\begin{figure}
\centering
\includegraphics[angle=270,scale=0.5]{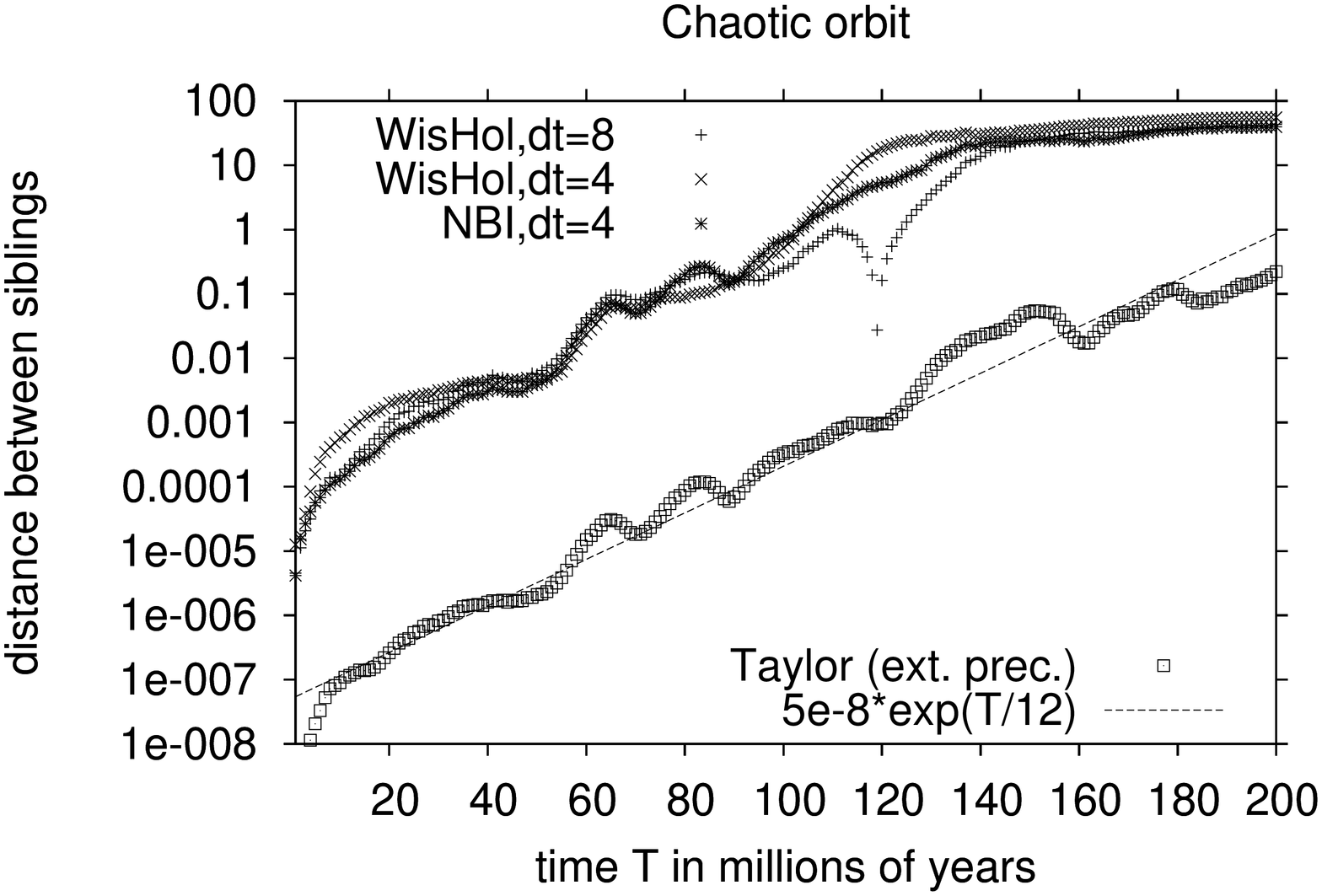}
\includegraphics[angle=270,scale=0.5]{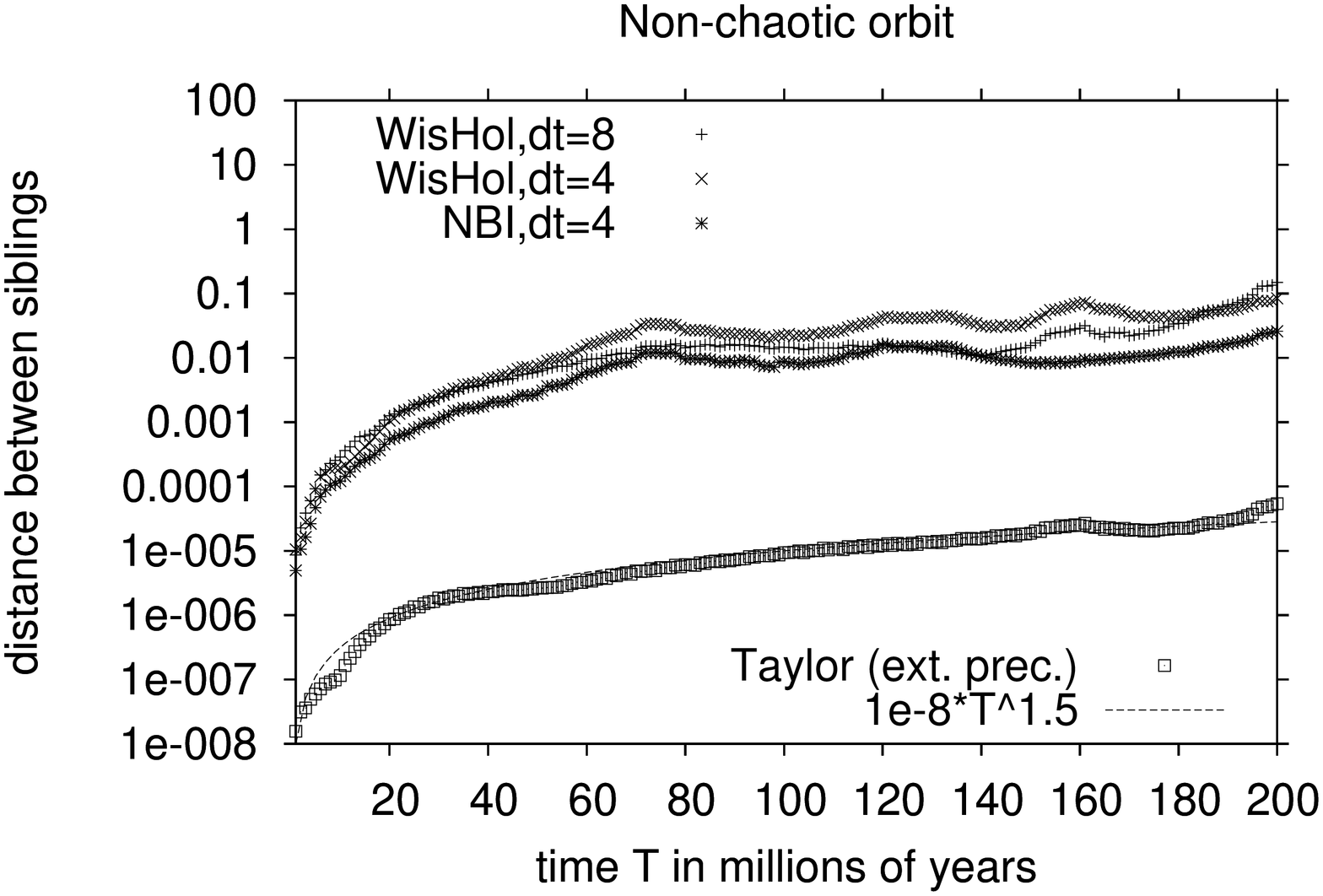}
\caption{Divergence between nearby trajectories, integrated with
four different numerical integrators.
Top figure: a chaotic trajectory with a Lyapunov time of about 12
million years.
Bottom figure: a trajectory showing no evidence of chaos over 200My.
Both trajectories are within observational uncertainty of
the outer planetary positions.}

\label{fig:surprise-all}
\end{figure}

\begin{figure}
\centering
\includegraphics[angle=0,scale=0.7]{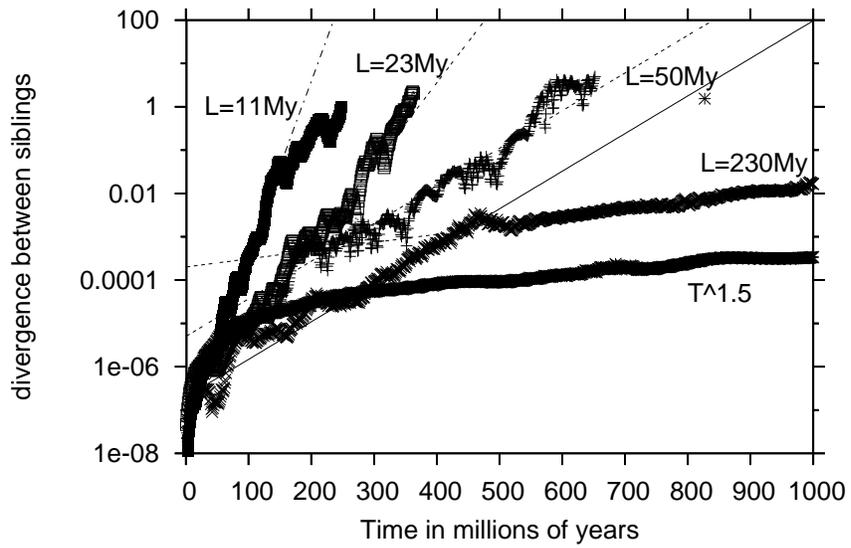}
\caption{Menagerie of Lyapunov times.
All trajectories originate within observational uncertainty,
although the regular one (labelled $T^{1.5}$, corresponding to
an infinite Lyapunov time) is the most ``authoritative'' initial condition
from JPL's DE405 ephemeris.
Finite ones range from 11My (million years) to 230My.
}
\label{fig:menagerie}
\end{figure}

Although it is impossible with a finite-time integration to demonstrate that
a trajectory is not chaotic, we abuse the term ``regular'' to mean
a trajectory that shows no evidence of chaos over timescales ranging from
200My (million years) to 1Gy ($10^9$ years).
Figure \ref{fig:surprise-all} plots the divergence between initial
conditions (ICs) that initially differ by an infinitesimal amount (1.5mm in
the semi-major axis of Uranus).  Chaos
manifests itself as exponential divergence between such ``siblings'',
while regularity is manifested by polynomial divergence.
Our results were consistent across three very different integration
schemes, and all schemes agreed with each other once the timesteps
were small enough to demonstrate convergence.
To estimate how the uncertainty volume is split between
chaotic and regular ICs, we chose 31 systems
within the observational uncertainty and integrated each for 200My.
We found that 21 of the 31 samples (about 70\%) were chaotic, and
10 (about 30\%) were regular.  When integrated for 1Gy, the percentage
of regular trajectories decreased to less than 10\%.
However, the
spectrum of observed Lyapunov times was enormous (Figure \ref{fig:menagerie}).
Furthermore, the most ``authoritative'' IC, the only one explicitly fit to
all observations in DE405\cite{Standish98-DE405}, shows no evidence of chaos after 1Gy.

\section*{Methods}

We use initial conditions (ICs) for the Sun and all 8 planets (excluding Pluto).
The inner planets are deleted, but their effect on the outer planets is crudely
accounted for by perturbing the Sun's position to the centre-of-mass of
the inner Solar System, and by augmenting the mass and momentum
of the Sun with the masses and momenta of the inner planets; this ensures\cite{MurrayHolman99}
that the positions of the chaotic zones are shifted by no more than about
one part in $10^{11}$.  The system
is then numerically integrated for $10^9$ years (1Gy) or until the
distance between siblings saturates, using
only Newtonian gravity.  The masses of all objects are held constant.
We also ignore many
physical effects which may be important to the detailed motion of
the planets\cite{Laskar99,VaradiRunnegarGhil03}.  However, 
we believe that none of these effects will alter the
chaotic nature of solutions.

Determining the existence of chaos depends critically on the quality of the
numerical integration scheme.  We ensure that our results are free
of numerical artifacts by performing our integrations using three
different algorithms.  First, we use the Wisdom-Holman symplectic mapping\cite{WisdomHolman91}
as implemented in the {\it Mercury 6.2} package\cite{ChambersMercury99},
with timesteps ranging from 400 days down to 2 days.
We find that using timesteps of 16 days or more lead to inconsistent results,
in that integrating the same IC with different timesteps results in significantly
different Lyapunov times \cite{WisdomHolman92}.  However, for a particular IC, results converge
to reliable Lyapunov times when the integration timestep is 8 days or less.
Our second integration scheme is the {\it NBI} package, which has been used to
demonstrate the non-existence of chaos in the outer Solar
System\cite{GrazierNewman96,GrazierNewmanKaulaHyman99,NewmanEtAlDDA2000}.
We use NBI's 14th-order Cowell-St\"{o}rmer integrator with modifications
by the UCLA research group led by William
Newman\cite{GrazierNewman95,GrazierNewman96,VaradiRunnegarGhil03}, with a timestep
of 4 days.
With these parameters, NBI is known to produce {\em exact} solutions to
double precision, at each timestep; one cannot do better than this
without maintaining machine precision while {\it increasing} the timestep (which
would lower the frequency at which the one-roundoff-per-step occurs).
Our third and highest precision integration scheme
is the {\it Taylor 1.4} package\cite{JorbaZou05}, which uses a 27th-order
Taylor series expansion and a 220-day timestep.
Arithmetic in {\it Taylor 1.4} is performed in 19-digit Intel Extended precision.
Like NBI, on each step {\it Taylor 1.4} produces results which are
exact to machine precision, except here the step is 220 days rather than
4 days, and the solution is exact to 19 digits rather than 16.
In the case of both {\it NBI} and {\it Taylor 1.4}, the numerical
error growth is effectively dominated by a single random roundoff
error per step.  We have verified the ``exact to machine precision per step''
properties of both {\it NBI} and {\it Taylor} by comparison
to quadruple-precision integrations which were locally accurate to 30 digits.
Once convergence was reached, all integrations had errors that grew
(in the absense of chaos) approximately as $t^{3/2}$, in accordance
with Brouwer's Law \cite{Brouwer37}.
Our 1Gy integrations using Taylor-1.4 conserved both energy and angular momentum
to almost 13 digits.  Numerical error caused the center of mass of the system to drift
by just 0.45km over 1Gy,
which is about 3 orders of magnitude smaller than the current observational error.
Thus, the numerical error in our 1Gy experiments is negligible compared to current observational error.

Figure \ref{fig:surprise-all} uses ICs
for the Sun and 8 planets as listed in the {\it Mercury 6.2}
package\cite{ChambersMercury99}, representing the positions of the planets
at JD2451000.5.  The semi-major axis of Uranus was increased by $2\times
10^{-6}$ AU (bottom figure), and $4\times 10^{-6}$ AU (top figure).
The percentage of ICs that are chaotic come from 31 eight-planet samples
from the latest JPL planetary ephemeris, DE405\cite{Standish98-DE405}.
We drew 21 sets of ICs starting at 9.5 Jan 1990 (JD2448235), separated
by 30-day intervals.  This samples ICs from a span of about 2 years,
so it includes samples such that all the inner planets are sampled at
widely varying positions in their orbits (before they are thrown into
the Sun).  We also drew 10 samples starting at the year 1900, separated
by 10-year intervals.  Figure \ref{fig:menagerie} uses ICs from all
of the above sources, and is not intended to represent a uniform sample.

Source code, initial conditions, and outputs are all available from the
author upon request.

\section*{References}
\bibliographystyle{nature}
\bibliography{nature}

\section*{Acknowledgements}
The author thanks Scott Tremaine, Norm Murray, Matt Holman, Kevin Grazier,
and Bill Newman for discussions, and Ferenc Varadi for source
code to the latest unpublished version of NBI.

\end{spacing}
\end{document}